\RequirePackage{fix-cm}
\documentclass[smallextended]{svjour3}       
\smartqed  
\usepackage{graphicx}
\usepackage{epsfig}
\usepackage{amsfonts}
\usepackage{amssymb}
\usepackage{amsmath}
\usepackage{bm}
\usepackage{hyperref}
\usepackage{mathrsfs}
\usepackage{subfigure}

\renewcommand\floatsep{6pt}
\renewcommand\intextsep{8pt}
\newcommand{\nb}{\nonumber}

\begin{document}

\title{Nonplanar On-shell Diagrams and Leading Singularities of Scattering Amplitudes
}

\author{Baoyi Chen, Gang Chen,  Yeuk-Kwan E. Cheung,  Yunxuan Li, Ruofei Xie, Yuan Xin}


\institute{
G. Chen \at Department of Physics, Zhejiang Normal University, Jinhua, Zhejiang Province, China\\
Department of Physics, Nanjing University, 22 Hankou Road, Nanjing 210093, P. R. China
 \and
           B.Y. Chen, Y.E. Cheung, Y.X.Li, R.F. Xie, Y. Xin \at
Department of Physics, Nanjing University, 22 Hankou Road, Nanjing 210093, P. R. China  
}

\date{Received: date / Accepted: date}

\maketitle

\begin{abstract}
Bipartite on-shell diagrams are the latest tool in constructing scattering amplitudes. In this paper we prove that a Britto-Cachazo-Feng-Witten (BCFW)-decomposable on-shell diagram process a rational top-form if and only if the algebraic ideal comprised of the geometrical constraints is shifted linearly during successive BCFW integrations.
With  a proper geometric interpretation of the constraints in the Grassmannian manifold, the rational top-form integration contours can thus be obtained, and understood,  in a straightforward way.
All rational top-form integrands of arbitrary higher loops leading singularities can therefore be derived  recursively, as long as the corresponding on-shell diagram is BCFW-decomposable.

\end{abstract}

\section{Introduction}
Scattering amplitudes are of profound importance in high energy physics describing the interactions of fundamental forces and elementary particles. The scattering amplitudes are widely studied for $\mathcal{N}=4$  super Yang-Mills theory and QCD.
At  tree level, BCFW recursion
relations~\cite{BCFW04-01,BCFW04-02,BCFW05,2012FrPhy...7..533F}
can be used to calculate  n-point amplitudes efficiently.
Unitarity cuts~\cite{Bern94,Bern95,Bern05} and generalized unitarity cuts%
~\cite{GUT04-01,GUT05-02,GUT05-06,Drummond:2008bq,Mastrolia:2012wf,Mastrolia:2012du,Mastrolia:2013kca,Multiloop2013}
combined  with  BCFW  for the rational terms work well at loop
level~\cite{Bern06,Carrasco11,Bern:2008ap,Eden1009,Eden1103,Eden11}.

Leading singularities~\cite{Cachazo2008}  are closely related to the unitarity cuts of loop-level amplitudes.
For planar diagrams in $\mathcal{N} = 4$ super Yang-Mills
\cite{Kanning2014,Ochirov2014gsa},  the leading singularities are invariant under Yangian symmetry~\cite{Broedel1403,Broedel2014,Beisert,Chicherin2014}, which is a symmetry combining conformal symmetry and dual conformal symmetry%
~\cite{Drummond:2008vq,Drummond:2008cr,Drummond:2009fd,Brandhuber:2009kh,NimaSimpleField}.
The leading singularity can also be used in constructing one-loop amplitudes by taking this as the rational coefficients of the scalar box integrals.
Extending this idea to higher loop amplitudes are reported in~\cite{GUT05-06,GUT0801,GUT0705,Spradlin:2008uu}.

A leading singularity can be viewed  as a contour integral over a Grassmannian manifold~\cite{NimaSMatrix,Paulos2014,BaiHe2014,Ferro2014gca,Franco2014csa,Elvang2014fja}.
This expression  of the leading singularity keeps many symmetries, in particular,  the Yangian symmetry, cyclic and parity symmetries, manifest.
On the one hand this new form makes the expression of amplitudes simple and hence easy to calculate. On the other hand it is related to the central ideas in algebraic geometry: Grassmmannian, stratification, algebraic varieties, toric geometry, and intersection theory etc..
For leading singularities of the planar amplitudes in $\mathcal{N}=4$~super Yang-Mills (SYM) ,  Arkani-Hamed et al~\cite{NimaGrass} proposed using positive Grassmannian to study them along  with  the  constructions of the bipartite on-shell--all internal legs are put on shell--diagrams~\cite{GeometryOnshell}.

Top-forms and the d``$log$'' forms of the Grassmannian integrals are systematically  studied  for planar diagrams.
Each on-shell  diagram corresponds  to a Yangian invariant, as shown in~\cite{Drummond:2009fd} at tree level and~\cite{Brandhuber:2009kh,NimaSimpleField,Elvang:2009ya} at loop level.
(See also~\cite{2003JHEP...10..017D,2004qts..conf..300D} for earlier works
and~\cite{2014arXiv1401.7274B,Frassek:2013xza,Amariti:2013ija,Bourjaily2013,CaronHuot:2011ky,Drummond:2010uq,Beisert2010,Drummond:2010qh,Feng2010,Alday:2010vh,Mason:2009qx,Beisert:2009cs,2009JHEP...06..045A,Bargheer:2009qu,2009JHEP...04..120A} for a sample of  interesting  developments thereafter,
and~\cite{Elvang:2013cua,Benincasa:2013faa,2012LMaPh..99....3B,Drummond:2011ic,Dixon:2011xs,Beisert:2010jq,Bartels:2011nz,2011JPhA...44S4011H,2011JPhA...44S4012B,Roiban:2010kk,Drummond:2010km,2005IJMPA..20.7189M,1993IJMPB...7.3517B} for a sample of  reviews and a new book~\cite{Henn:2014yza}.)

We report, in this paper, our detailed and systematic studies of the nonplanar on-shell  diagrams which can be decomposed in by removing BCFW-bridges  and applying U(1) decoupling relation of the  four- and three-point amplitudes (or just decomposable diagrams for convenience). For  wide classes of  leading singularities, the corresponding on-shell diagrams are decomposable diagrams.  We first  construct   the chain of  BCFW-decompositions  for  the on-shell diagrams. During this process we obtain the unglued diagram by cutting an internal line. We prove any unglued diagrams can be categorized into three distinct classes which can be subsequently turned into identity utilizing crucially the permutation relation of generalized Yangian Invariants~\cite{Du:2014jwa}.  This construction is presented in Section~\ref{sec:BCFWChain}.

We then proceed to study the geometry of the leading singularities. We are interested in the constraints encoded in the Grassmannian manifolds and how these constraints determine the integration contours in the top-forms. As the cyclic order is destroyed by non-planarity the integrand of Grassmannian integral also needs to be constructed from scratch. To achieve the above goals we attach non-adjacent BCFW-bridges to the planar diagrams and observe how the integrands and the C-matrices transform.
Further we can construct the (rational) top-form including both integrand and integration contour of any nonplanar leading singularity by attaching (linear) BCFW-bridges to the identity diagram in the reverse order of the BCFW decomposition chain from the previous section. This construction is presented in Section~\ref{sec:TopForm}

\section{Scattering amplitudes: BCFW decomposition}
\label{sec:BCFWChain}
\begin{figure}[htbp]
  \centering
  \includegraphics[width=0.6\textwidth]{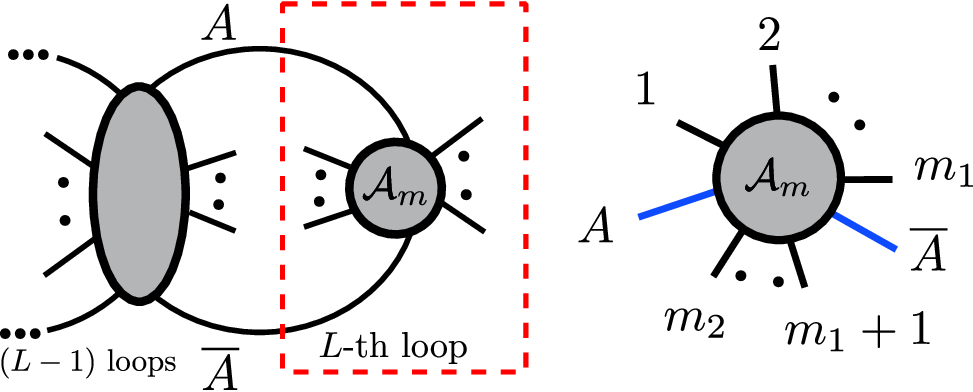}\\
  \caption{The  L-loop's d$log$ form can be obtained by reducing to (L-1)-loop problem. }
  \label{fig:NLoopCut}
\end{figure}

In an on-shell diagram representing an $L$-loop leading singularity, we are free to pull out a planar  sub-diagram (unglued diagram) between two internal loop lines--both are also on-shell as shown in Fig.~\ref{fig:NLoopCut}
\footnote{%
In this paper we assume there are more than two external lines for this planar sub-diagram. If there is only one external line in the sub-diagram, we need to use other method which is presented in our following paper.}. Locally the sub-diagram is planar except that we cannot perform BCFW integrations on these two loop lines. We proved that every such sub-diagram, upon the removal of all BCFW bridges in the permutations, can be  casted  into one of the three distinct types of skeleton graphs. $U(1)$-decoupling relation can be further performed on the latter two types. And the $L$-th loop is unfolded. Unfolding the loops recursively, we  obtain the BCFW decomposition chain for the  leading singularities  of any $L$-loop nonplanar amplitudes. In other words the BCFW chain captures all the information of the leading singularities of the L-loop nonplanar graphs. We are thus able to reconstruct the on-shell diagrams by attaching BCFW-bridges from the identity.

In this section, we will introduce a systematic way of finding the BCFW bridge decomposition chain from the marked permutations\footnote{%
Marked permutations refer to permutations with two end  points treated specially--the two points are allowed to marked to themselves or to each other.}
 of the  unglued diagrams

\subsection{From permutations to BCFW decompositions}

In this subsection, we derive the BCFW-bridge decomposition chain from the permutation of the unglued diagram. First we perform BCFW bridge decompositions on the unglued diagram according to marked permutation, leaving the cut lines untouched. Upon the removal of all adjacent bridges, we will arrive at three categories of skeleton diagrams, noticing that the category type is invariant under the BCFW bridge decomposition. Next for each category of skeleton diagram we construct a specific recipe to decompose it to identity.

\begin{figure}[htbp]
  \centering
  \includegraphics[width=0.4\textwidth]{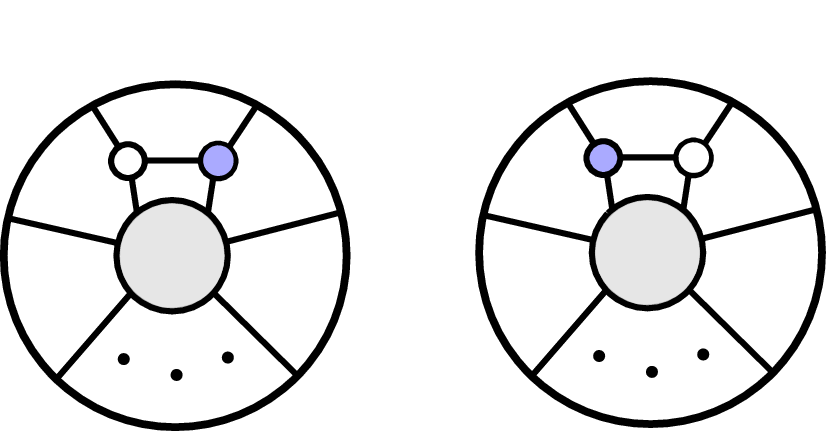}
  \caption{white-black bridges and black-white bridges}\label{fig:BCFWbridge}
\end{figure}

\paragraph{From unglued diagram to skeleton diagram}
All unglued diagrams can be categorized into three groups depending on the permutations of the two cut lines, denoted as  $A$ and $\bar{A}$,
\begin{itemize}
\item[~]
\begin{itemize}
\item[\textbf{(1)}] $\sigma_1(A) \neq \bar{A} ~\&\&~ \sigma_1(\bar{A}) \neq A $;
\item[\textbf{(2)}] $\sigma_2^a(A) = \bar{A} ~\&\&~\sigma_2^a(\bar{A}) \neq A $, or $\sigma_2^b(\bar{A}) = A ~\&\&~\sigma_2^b(A) \neq \bar{A}$;
\item[\textbf{(3)}] $\sigma_3(A) = \bar{A} ~\&\&~\sigma_3(\bar{A}) = A$.
\end{itemize}
\end{itemize}
To decompose an unglued diagram, the first step is the full removal of two types of adjacent bridges on the target diagram:
 the \textit{white-black bridge} and the \textit{black-white bridge}
  as shown in Fig.\ref{fig:BCFWbridge}.
The changes to the permutation after removing either of them are, respectively,
 $\sigma\rightarrow \sigma'=Z_2(k, k+1)\cdot\sigma$ and $\sigma\rightarrow \sigma'=Z_2(\sigma^{-1}(k),\sigma^{-1}( k+1))\cdot\sigma$, where $Z_2(k, k+1)$ is a $Z_2$ permutation between line $k$ and $k+1$\cite{NimaGrass}.
For an unglued diagram arisen from a nonplanar leading singularity, the cut line should not be involved in BCFW bridge decompositions as the pair of marked lines are to be glued back eventually. Thus we should restrict the set of allowed BCFW bridge decompositions to those preserving the two marked legs. By following this restriction, the group our target unglued diagram originally belongs to will not alter during bridge decompositions.

\begin{figure}[htbp]
  \centering
  \subfigure[External line Pair]{
  \label{fig:clean:clean1}
  \includegraphics[width=0.3\textwidth]{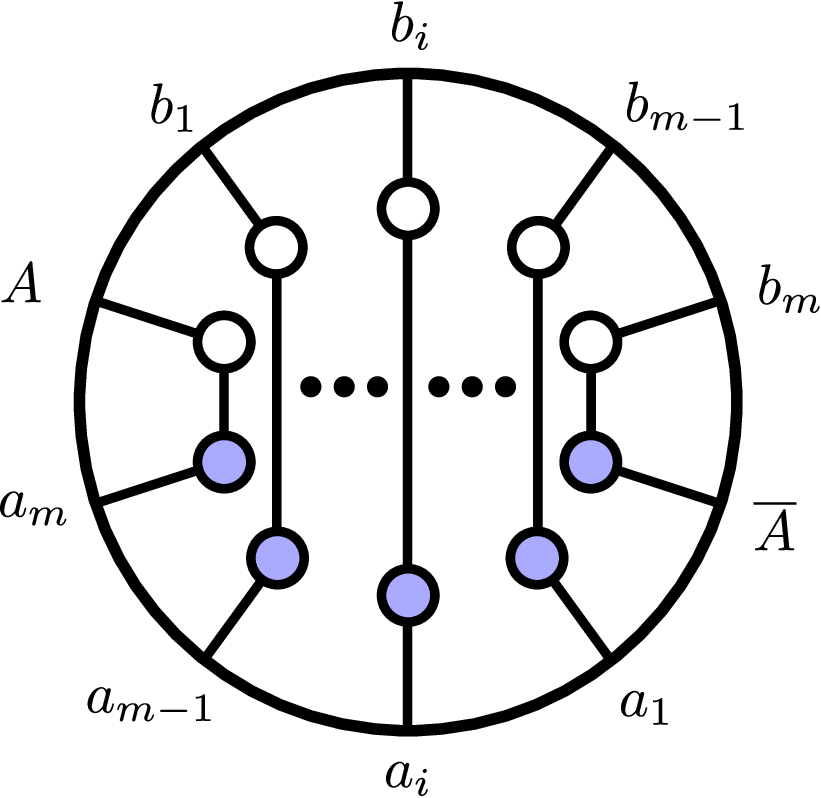}
  }
  \subfigure[Black-White Chain]{
  \label{fig:clean:clean2}
  \includegraphics[width=0.3\textwidth]{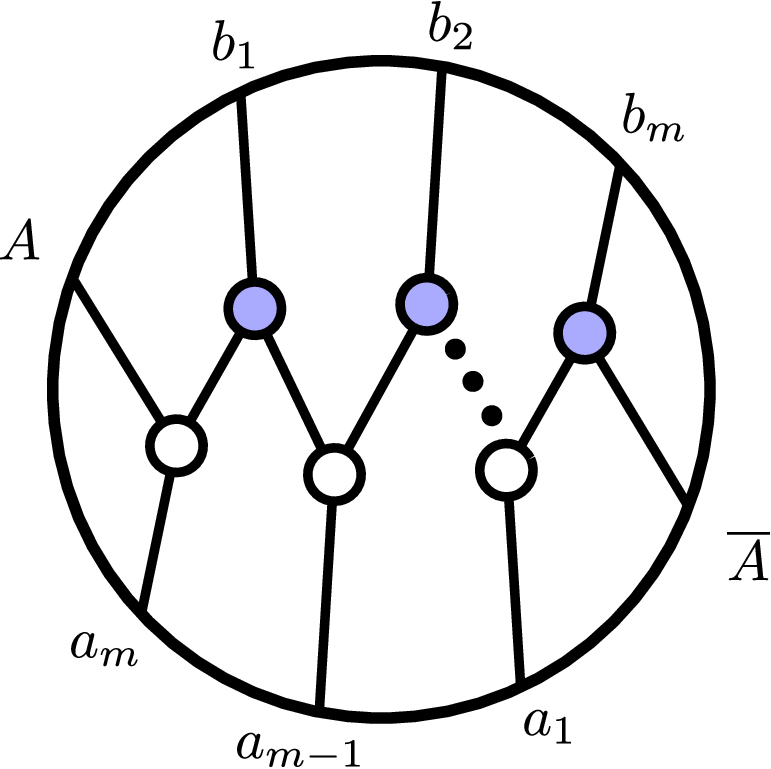}
  }
  \subfigure[Box Chain]{
  \label{fig:clean:clean3}
  \includegraphics[width=0.3\textwidth]{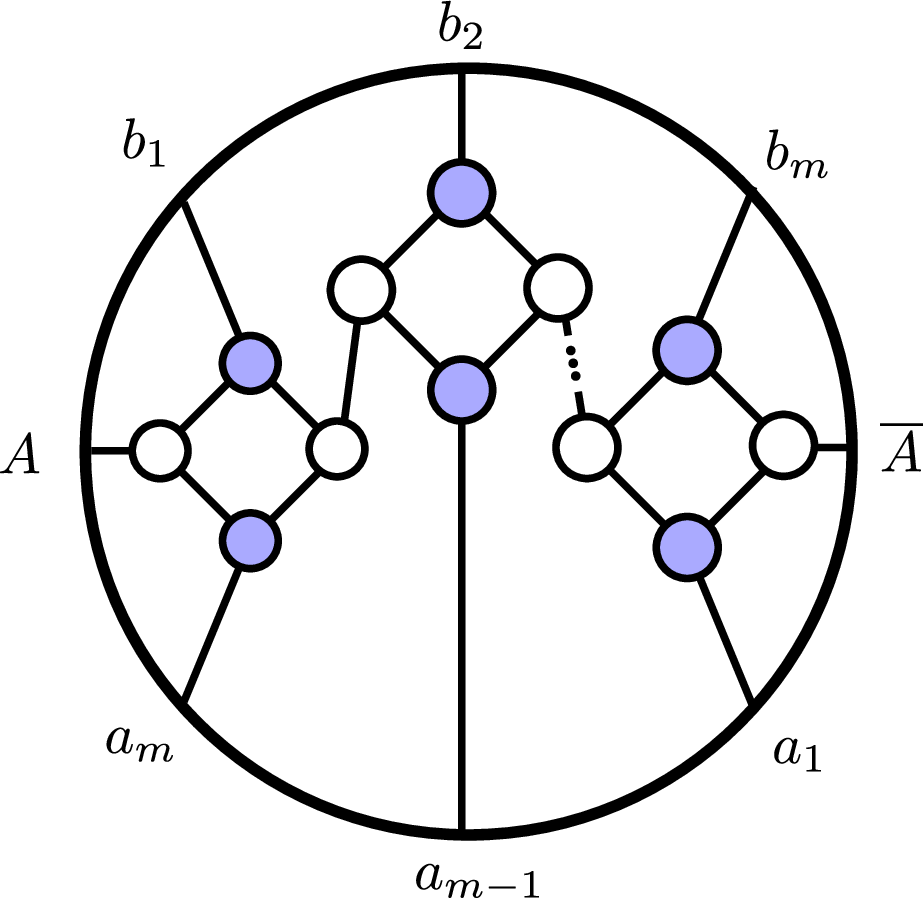}
  }
  \centering
  \caption{Skeleton Diagrams}
  \label{fig:clean}
\end{figure}

Due to the existence of nonadjacent bridges, an unglued diagram cannot be fully decomposed and will pause at a certain diagram. It is easy to see that after removing all BW- and WB- bridges the three groups of unglued diagrams will fall into \textit{External line pair}, \textit{Black-White Chain} and \textit{Box Chain} respectively. The three categories are named after their general patterns as shown in Fig.~\ref{fig:clean}. We refer to any diagram belonging to the above three categories as the \textit{``skeleton diagram''}, naming after its skinny looks. As long as we can fully decompose all three skeleton diagrams, it is then direct to obtain the complete decomposition chain of any unglued diagram.

\paragraph{From skeleton diagram to identity}
\begin{figure}[htbp]
  \centering
  \includegraphics[width=0.8\textwidth]{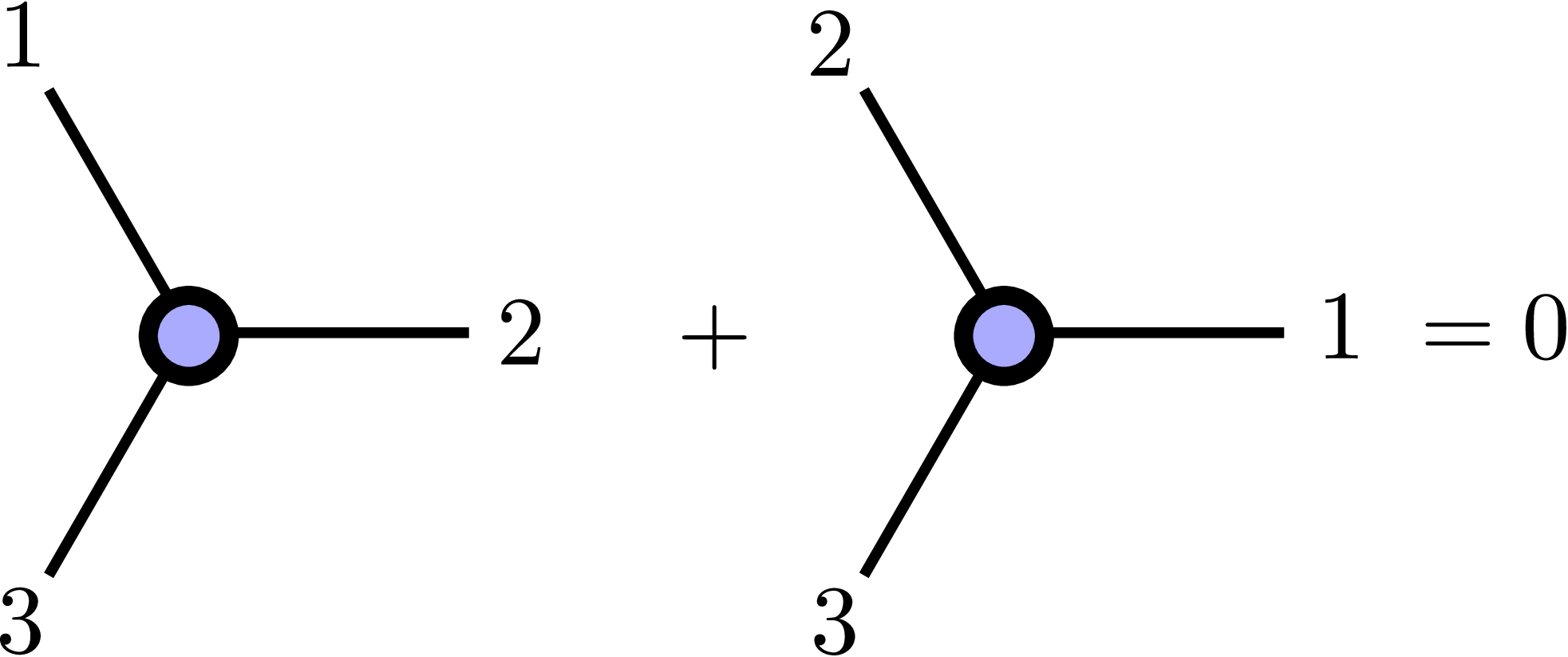}
  \caption{$U(1)$ decoupling relation of 3-point amplitudes.}
  \label{fig:relation3}
\end{figure}
\begin{figure}[htbp]
  \centering
  \includegraphics[width=0.8\textwidth]{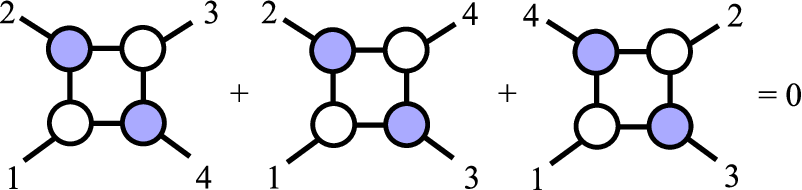}\\
  \caption{$U(1)$ decoupling relation of 4-point amplitudes.}
  \label{fig:U1relation}
\end{figure}
\begin{itemize}
\item \noindent{\bf External Line Pair:} Most external lines are paired. The external lines next to the internal cut line may also attach to the black/white vertices or be paired with the internal cut line as shown in Fig~\ref{fig:clean:clean1}. For this type of on-shell diagrams, gluing back the internal lines and removing all pairs will lead to the identity.

\item \noindent{\bf Black-White Chain:} In this case, white and black vertices are connected together recursively, as shown in Fig~\ref{fig:clean:clean2}.
To further decompose, we use the amplitudes relation $\mathcal{A}(a_1, a_2, a_3)=-\mathcal{A}(a_1,a_3,a_2)$ (see Fig~\ref{fig:relation3}) to twist one down-leg to up-leg. Then an adjacent bridge will appear. By removing the new appeared BCFW bridge, the diagram is unfolded into a planar diagram. The diagram can then be decomposed to identity according to its permutation~\cite{NimaGrass}.

\item \noindent{\bf Box Chain:} In this case, the diagram is composed by boxes linked to a chain, as shown in Fig~\ref{fig:clean:clean3}. Using the $U(1)$ decoupling relation~\cite{KK} of the four point amplitudes (see Fig~\ref{fig:U1relation}), this diagram turns into the sum of two diagrams with adjacent BCFW bridges.
The non-adjacent legs of the box will become adjacent under this operation. Performing adjacent BCFW decompositions on both diagrams will unfold the loop and arrive at two planar diagrams, which can be decomposed to identity.
\end{itemize}

\section{Scattering amplitudes: the \textit{Top--form}}
\label{sec:TopForm}
Through   the BCFW bridge decompositions  we obtain the d$log$ form characterized  by the bridge parameters.
The d$log$ form can be viewed as an explicit parameterization of a more general integration over the Grassmannian manifold, which is invariant under the $GL(k)$ transformations.  The invariant form, known as the ``top-form,''  for planar diagrams has been  constructed in \cite{NimaGrass}. In this section, we  construct the top-form for the  nonplanar leading singularities. Recent progress on nonplanar on-shell diagram can be  find in \cite{MHVNP,franco2015non}

For planar diagrams, the top-form manifests the Yangian symmetry:
the leading singularities can be written as multidimensional residues in the  Grassmannian manifold $\mathcal{G}(k,n)$,
\begin{equation}
\mathcal{T}_n^k=\oint\limits_{C\subset\Gamma}\frac{d^{k\times n}C}{vol(GL(k))}\frac{\delta^{k\times 4}(C\cdot\widetilde{\eta})}{f(C)}\delta^{k\times 2}(C\cdot\widetilde{\lambda})\delta^{2\times(n-k)}(\lambda\cdot C_{\perp}^{\top}),
\label{eq:topform}
\end{equation}
where $\Gamma$ is a sub-manifold of $\mathcal{G}(k,n)$.
$\Gamma$ is constrained by  a set of  linear relations among the columns of $C$--certain minors of $C$ be zero.
As  any  function of the minors of $C$,  $f(C)$,  has the scaling property $f(tC)=t^{k\times n}f(C)$.

To construct  the top-form for nonplanar leading singularities, we need to determine the integration contour $\Gamma$ and the integrand $f(C)$.
Since the integration contour is constrained by a set of geometrical relations linear in $\alpha$'s, we make use of  the BCFW chain  we obtained in Section~\ref{sec:BCFWChain} to look for  all geometric  constraints, fixing  $\Gamma$ in the process.
Next we will see,  with the BCFW approach extended to loop-level, the integrand of  the top-form can be calculated by  attaching BCFW bridges.

\subsection{Geometry and the  BCFW-Bridge Decomposition}
\label{topFormBridge}
In this subsection  we shall introduce the method of searching for  geometric  constraints in the Grassmannian matrix. Geometric constraints are linear relations among columns of $C$ matrix. In fact, the total space is taken as $(k-1)$-dimension projective space. Each column denoted by the index of the external line can be map to a point in the projective space. Each time we attach a bridge  a constraint will be fixed and the geometry constraints  change accordingly. In the Grassmannian matrix, adding a white-black bridge on external lines $\mathbf{a}$, $\mathbf{b}$ yields a linear transformations of the two columns,
$\mathbf{a}$ and $\mathbf{b}$,  $\mathbf{a} \longrightarrow \widehat{\mathbf{a}} =\alpha \mathbf{a}+\mathbf{b}$;
whereas  adding a black-white bridge means $\mathbf{b}\longrightarrow\widehat{\mathbf{b}}=\alpha \mathbf{b}+\mathbf{a}$.

For convenience, we divide the geometric constraint into two types: Simple coplanar constraint and tangled coplanar constraint. Simple coplanarity is just the coplanarity among the points corresponding to the external line. For the tangled coplanar constraint, at least one point is formed by the intersection of super-planes characterized by the point of the external line. We first present an example for each case. 

\begin{figure}[htbp]
  \centering
  \includegraphics[width=0.5\textwidth]{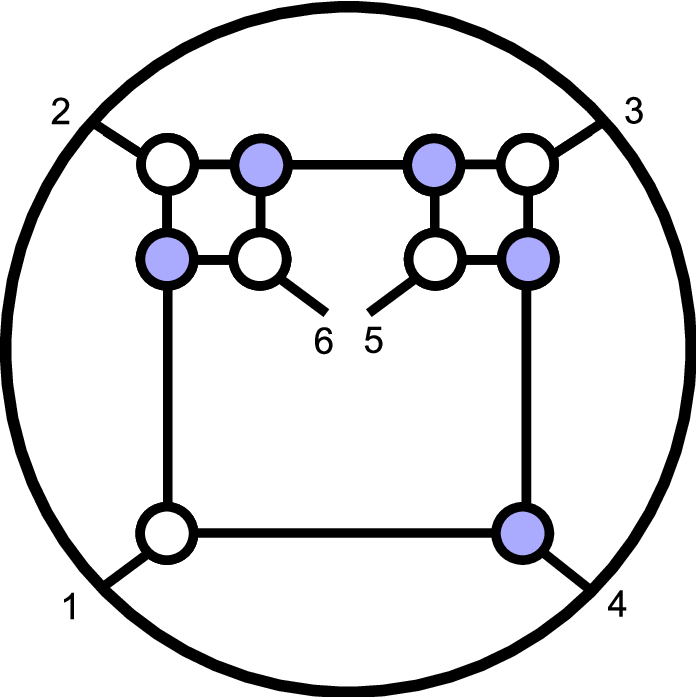}\\
  \caption{The BCFW bridge to open the loop of the diagram.}
  \label{fig:geometry6point}
\end{figure}

\paragraph{Example for simple coplanarity}As an explicit example, we work out $\mathcal{A}_{6}^{3}$'s geometry shown in Fig.~\ref{fig:geometry6point}. This diagram becomes planar upon removal of a white-black bridge ($1$,$4$). The remaining BCFW bridge decomposition is:
$$(1,2)\rightarrow (2,3)\rightarrow (3,4)\rightarrow (2,3)\rightarrow (1,2)\rightarrow (3,5)\rightarrow (2,6)$$
Linear relations in identity are then:
$$(4)^0,\ (5)^0,\ (6)^{0}$$
In the Grassmannian matrix, all elements in the three columns are zero. We then reconstruct the diagram through attaching BCFW bridges.
\begin{table}[htp]
\label{SimpleBridge}
\caption{The evolution of the geometry constraints under adding BCFW bridges. The first row is the linear relation in the identity diagram and the column on the left represents the bridge decomposition chain.}
\begin{center}
\begin{tabular}{|c|c|}
\hline
Bridge &coplanar constraints \\ \hline
 \textit{begin}&$(4)^{0}$\ $(5)^{0}$\ $(6)^{0}$ \\ \hline
  (2,6) & $(4)^{0}$\ $(5)^{0}$\ $(2,6)^{1}$\\  \hline
  (3,5) & $(4)^{0}$\ $(3,5)^1$\ $(2,6)^{1}$\\  \hline
  (1,2) & $(4)^{0}$\ $(3,5)^1$\ $(1,2,6)^{2}$\\ \hline
  (2,3) & $(4)^{0}$\ $(2,3,5)^2$\ $(1,2,6)^{2}$\\  \hline
  (3,4) & $(3,4)^{1}$\ $(2,(3,4)^1,5)^2$\ $(1,2,6)^{2}$ \\ \hline
  (2,3) & $(2,3,4,5)^2$\ $(1,2,6)^{2}$ \\  \hline
  (1,2) & $(3,4,5)^2$\ $(1,2,6)^{2}$ \\  \hline
  (1,4) & $(1,2,6)^{2}$ \\
  \hline
\end{tabular}
\end{center}
\end{table}
There are eight bridges needed to construct the nonplanar diagram. Each step will diminish one coplanar  relation. For instance, the first step is adding a white-black bridge on external line $2$ and $6$, leaving $column\ 6$ to become $c_6+\alpha c_2$. The relation $(6)^{0}$ then becomes $(2,6)^{1}$. Similarly upon attaching bridges $(3,5), (1,2), (2,3)$, the coplanar constraints are  $(4)^{0}$, $(2,3,5)^2$, $(1, 2, 6)^{2}$. Upon attaching bridge $(3,4)$, constraint $(4)^{0}$ becomes $(3,4)^{1}$. This means that point $3$ and $4$ merge to a one point. Then the  constraint $(2,3,5)^2$ can be written as  $(2,(3,4)^{1},5)^2$. Attaching the bridges consecutively as shown in Tab. \ref{SimpleBridge}, we can finally get the coplanar constraint $(1,2,6)^{2}$ for the on-shell diagram. 

\begin{figure}[htp]
 \centering
\includegraphics[width=0.5\textwidth]{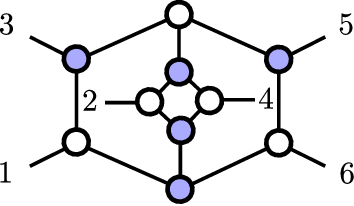}~~~~~~~
\caption{An example of  nonplanar two loop diagram $\mathcal{A}_{6}^{3}$. }
\label{fig:A63twoloop}
\end{figure}

\paragraph{Example for tangled coplanarity}We consider a nonplanar 2-loop diagram, $\mathcal{A}_{6}^{3}$ as shown in Fig.~\ref{fig:A63twoloop}.
The BCFW decomposition chain is
\begin{eqnarray}
(1,3)\rightarrow(3,5)\rightarrow(2,3)\rightarrow(1,2)\rightarrow(3,4)\rightarrow(2,3)\rightarrow(3,5)\rightarrow(3,6).
\end{eqnarray}
Then, the top-form of the diagram can be reconstructed by attaching these bridges one by one. There are eight bridges and each one diminish a coplanar constraint as shown in Tab.~\ref{TwoLoopGeo}
\begin{table}[htp]
\label{TwoLoopGeo}
\caption{The evolution of the geometry constraints under adding BCFW bridges. }
\begin{center}
\begin{tabular}{|c|c|}
\hline
Bridge &coplanar constraints \\ \hline
\textit{begin} &$(4)^{0}$\ $(5)^{0}$\ $(6)^{0}$ \\ \hline
  (3,6) & $(4)^{0}$\ $(5)^{0}$\ $(3,6)^{1}$\\  \hline
  (3,5) & $(4)^{0}$\ $(3,5)^1$ \ $(5,6)^1$\\  \hline
  (2,3) & $(4)^{0}$\ $(2,3,5)^2$\ $(5,6)^{1}$\\ \hline
  (3,4) & $(34)^{1}$\ $(2,3,5)^2$\ $(5,6)^{1}$\\  \hline
  (1,2) & $(3,4)^{1}$\ $(5,6)^{1}$ \\ \hline
  (2,3) & $(2,3,4)^2$\ $(5,6)^{1}$ \\  \hline
  (3,5) & $(2,3,4)^2$\ $(3,5,6)^{2}$ \\  \hline
  (1,3) & ${(234)\over(214)}-{(356)\over(156)}=0$ \\
  \hline
\end{tabular}
\end{center}
\end{table}

\begin{figure}[htp]
\centering
\includegraphics[width=0.5\textwidth]{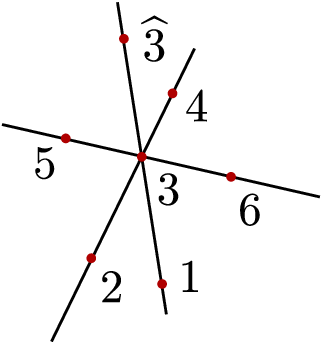}
\caption{The constraint of the diagram is $\big((1\,\widehat 3)\cap(2\,4)\,5\,6\big)$. In projective space, it means line $(1\,\widehat 3)$, line $(2\,4)$ and line $(5\,6)$ all intersect on point $3$. Such a constraint can only appear in multi-loop nonplanar on-shell diagrams.}
\label{GeoA63}
\end{figure}

The  first seven bridges attached yield simple coplanar constraints. Then the geometry constraints is $(2,3,4)^2 , (3,5,6)^{2}$, which indicates that points $2,3,4$ and $3,5,6$ are collinear respectively as shown in Fig. \ref{GeoA63}.  Then we attach the last bridge. As discussed above, point $3$ is shifted to $\hat 3$ along the line-(31). It seems the $2,{\hat 3},4$ and $\hat{3},5,6$ are not collinear anymore and the two constraints are removed together. In fact, according to Fig. \ref {GeoA63}, there is another coplanar constraint that is the intersect point $(1\,\widehat 3)\cap(2\,4)$ lie in the line of $(5\,6)$.  For convenience, we denote this tangled coplanar relation as $\big((1\,\widehat 3)\cap(2\,4)\,5\,6\big)$.The geometry evolution under the last bridge $(1,3)$ is shown in Fig.~\ref{GeoA63}.

\paragraph{General simple coplanar constraints}We first discuss  the cases without tangled coplanar constraints. We can classify the coplanar constraints into four sets according to the elements:
$$\begin{array}{l}
1.~(\mathbf{a}, \, \mathbf{b},1,2...m_{1})^{k_1}\\
2.~(\mathbf{a},1,2...m_{2})^{k_2}\\
3.~(\mathbf{b}, 1,2...m_{3})^{k_3}\\
4.~(1,2,...m_{4})^{k_4}
\end{array}$$
where $k_1$ to $k_4$ are the ranks of the minors.
Without loss of generality,  we shall make $k_1<m_{1}+2$, $k_2<m_2+1$, $k_3<m_3+1$ and $k_4<m_4$.
We call the minor \textit{complete} if only if adding any other column to the matrix  will make the rank increase by  one.
From now on  we shall assume that all the minors in the above sets are complete in the following discussion.
In fact,  incomplete minors can always be transformed into the complete ones by adding  to the bracket all the necessary  elements while keeping the rank unaltered.

Attaching a white-black bridge  does not change the rank of minors in $set~1$ since $a$ and $b$ are both in this set.
The minors in $set~2$ and $set~4$ remain unaltered since $b$ is excluded from these two sets.
$(\alpha \mathbf{a} + \mathbf{b})\notin span\{\mathbf{b},\mathbf{1},\mathbf{2}...\mathbf{m_{3}}\}$,
thus after adding a white-black bridge the only set with its rank altered is $set\ 3$.
The minors in $set\ 3$ can generate two new linear relations: $(a,b,1,2...m_{3})^{k_3+1}$ and $(1,2...m_{3})^{k_3}$.
Similarly, upon attaching a black-white bridge, the minors in $set\ 2$ will become $(a,b,1,2...m_{2})^{k_2+1}$ and $(1,2...m_{2})^{k_2}$.
We have completed the  discussion of  how constraints alter during each step of bridge decompositions.

Next we turn to attaching bridges starting from the identity with the identity diagram being
a matrix with  $n-k$  columns  of zero vectors.
Each time we attach a BCFW bridge, the number of independent geometric  constraints will decrease by  one.
This can be proved through the following procedure.
Attaching the bridge ($\mathrm{a}$,\, $\mathrm{b}$) affects the linear relation involving  $\mathbf{b}$.
The only exceptions are the relations containing both $a$ and $b$, which will not be affected by the bridge  ($\mathrm{a}$,\, $\mathrm{b}$).
$$
(b,1,2...m)^{k}  \longrightarrow \left\{
                          \begin{array}{l}
                            (\mathbf{a},  \mathbf{b}  ,(1,2... \mathbf{m})^k )^{k+1} \\
                            (1,2...m)^{k} \\
                          \end{array}\right.
$$
If $k=m$, the linear relation $(1,2...m)^{k}$ does not give rise to any constraint,  thus $(a,b,1,2...m)^{k+1}$ has one higher rank than $(b,1,2...m)^{k}$. The constraints' number is then diminished by one upon attaching the bridge. If $k<m$, the coplanar constraints $(b,1,2...m)^{k}$ can be decomposed to $(b,(1,2...m)^{k})^{k}$ and $(1,2...m)^{k}$. The independent constraints after attaching the bridge are $(a,b,(1,2...m)^{k})^{k+1}$ and $(1,2...m)^{k}$. Comparing the constraints between $(b,(1,2...m)^{k})^{k}$ and $(a,b,(1,2...m)^{k})^{k+1}$, the number of constraints is reduced by one upon adding the BCFW bridge.

\paragraph{General Tangled Coplanar Constraints}
Now we discuss the  tangled coplanar constraints. When we attach a BCFW bridge, $X \rightarrow \widehat{X} =X + \alpha Y$\footnote{The bridge at least shifts one of the constraints linearly. We will show in next section that this condition is equivalent to that the top-form is rational.}.  There are  two constraints $(XA_1A_2\ldots A_{i_1})^{i_1}$ and $(XB_1B_2\ldots B_{i_2})^{i_2}$
both containing the shifting leg. Without losing of generality, we assume that  there is no constraints among $A_1, A_2\ldots A_{i_1}$ and $B_1, B_2\ldots B_{i_2}$. 
Upon attaching a BCFW bridge a tangled constraint could be obtained:
\begin{equation}
\label{eq:interRelation}
\alpha=\frac{V^{i_1+1}(\widehat{X}A_1\ldots A_{i_1})}{V^{i_1+1}(YA_1\ldots A_{i_1})}=\frac{V^{i_2+1}(\widehat{X}B_1\ldots B_{i_2})}{V^{i_2+1}(YB_1\ldots B_{i_2})},
\end{equation}
where $V$ stands for the volume of the hyperpolyhedron.
\begin{figure}[htp]
\centering
    \includegraphics[width=0.35\linewidth]{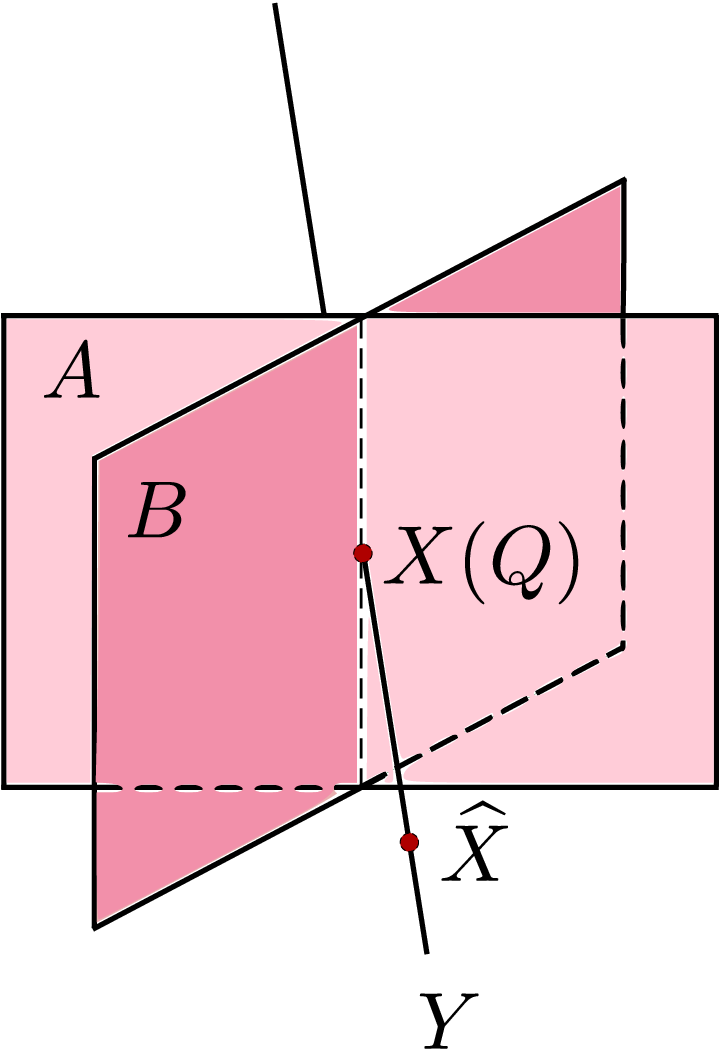}
\hspace{20pt}
    \includegraphics[width=0.45\linewidth]{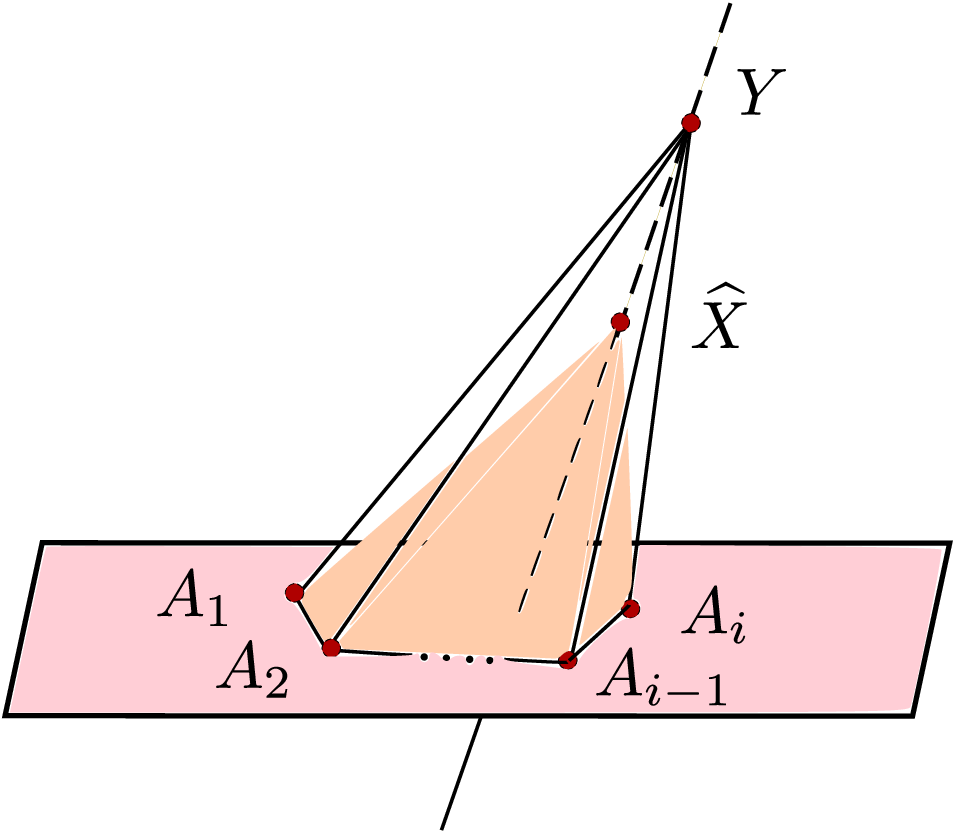}
\setlength{\abovecaptionskip}{5pt}
\caption{The diagrammatic notation of Eq.~\ref{eq:geo2alg}. The plane or hyperplane $A=Span(XA_1A_2\ldots A_i)$ and $B=Span(XB_1B_2\ldots B_i)$. $\alpha$ can be regarded as the ratio of two hyperpolyhedron's volumes.}
\label{fig:Intersect}
\end{figure}
It is indeed a geometric constraint on the Grassmannian manifold--the point of intersection of one line and  one hyperplane  lying on  another hyperplane:
\vspace{-5pt}
\begin{equation}
\Big((\widehat{X}Y)\cap(B_1B_2\ldots B_{i_2})A_1A_2\ldots A_{i_1}\Big).
\label{eq:geo2alg}
\vspace{-5pt}
\end{equation}
We denote the intersection point of
$(\widehat{X}Y)\cap(B_1B_2\ldots B_i)$ as $Q=C_1\widehat{X}+C_2Y$.
Since $Q$ also lies in the plane $(B_1B_2\ldots B_i)$, 
$$C_1V^{i_2+1}(\widehat{X}B_1B_2\ldots
B_{i_2})+C_2V^{i_2+1}(YB_1B_2\ldots B_{i_2})=0.$$
Thus the initial constraint
$(QA_1A_2\ldots A_{i_1})^{i_1}$ directly yields Eq.~\ref{eq:interRelation}.
One may have noticed that the point $Q$ is precisely the point $X$ before shifting.
If we go on attaching  another bridge that involves
this  tangled constraint, the set of constraints can again  be written as minors of the Grassmannian matrix.

Therefore we conclude that general constraints can always be labelled using nested spans and intersections.
Consider  attaching  a linear BCFW bridge $(Y,X)$ in an arbitrary amplitude, a constraint to be shifted is
\begin{small}
\begin{eqnarray}
     \label{eqn:GL}
M(X)&\equiv&\Big(\cdots (XA^{(0)}_1 A^{(0)}_2\cdots A^{(0)}_{a_0}) \cap (B^{(1)}_1 B^{(1)}_2\cdots B^{(1)}_{b_1})A^{(1)}_1 A^{(1)}_2\cdots A^{(1)}_{a_1})\nb\\
 && \cap (B^{(2)}_1 B^{(2)}_2\cdots B^{(2)}_{b_2})A^{(2)}_1 A^{(2)}_1\cdots A^{(2)}_{a_2}) \nb \cdots \cap (B^{(m)}_1 B^{(m)}_2\cdots B^{(m)}_{b_m}) A^{(m)}_{1} A^{(m)}_2\cdots A^{(m)}_{a_m}\Big)~,
\end{eqnarray}
\end{small}
with  $X$ being the external line to be shifted, $A^{(\cdot)}$ and $B^{(\cdot)}$ denoting
two sets of external lines.
If column $X\in Set[A^{(\cdot)}]$ or $X\in Set[B^{(\cdot)}]$, they can be freely replaced by $\hat X$ after attaching the bridge involving $X$.
Otherwise the constraint will be a nonlinear function of 
$\alpha$, resulting  in an irrational top-form.
We present a counter example
in Appendix 
to illustrate this point.
We can then simplify  $M(X)$ as follows
\vspace{-5pt}
\begin{equation}
\Big(L^{(m)}(X)\cap(B^{(m)}_1 B^{(m)}_2\cdots B^{(m)}_{b_m}) A^{(m)}_{1}\bar A^{(m)}_2\cdots \bar A^{(m)}_{\bar a_m}\Big),\nonumber\vspace{-5pt}
\end{equation}
where $\bar A^{(\cdot)}$ are some points or hyperplanes composed by $A^{(\cdot)}$ and  $B^{(\cdot)}$ and are easily obtained through a simple relation,
$$ (A_1 A_2 A_3)\cap P
= \left(\left((A_1 A_2)\cap P\right)\left((A_2 A_3)\cap P\right)\right)
= \left(((A_1 A_2)\cap P)\bar A\right).$$
After the shift the constraint $M(X)$ is removed.
In order to obtain the other constraints, we look for
the $\widehat C$ representation of $X$.
This is achieved by unfolding  the nested intersections level by level.
\begin{figure}
  \centering
  \includegraphics[width=0.35\columnwidth]{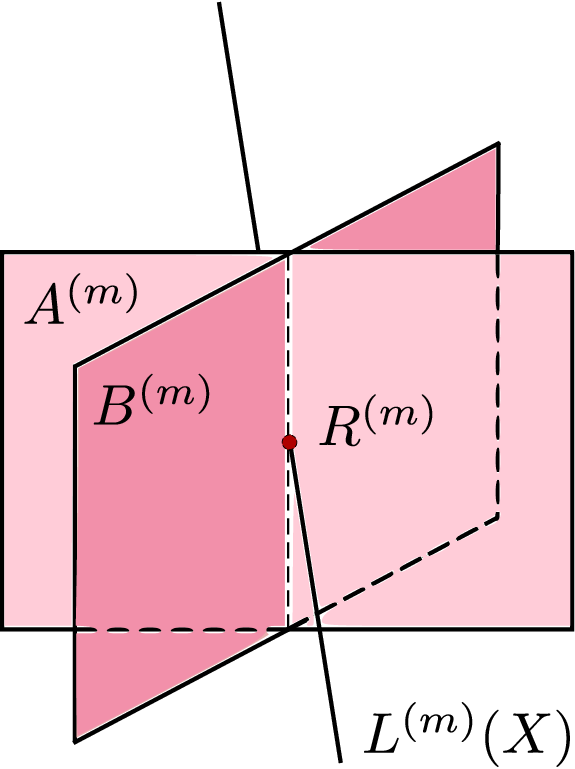}
  \hspace{20pt} \raisebox{30pt}{\Large$\rightarrow$} \hspace{20pt}
  \raisebox{-12pt}{
  \includegraphics[width=0.35\columnwidth]{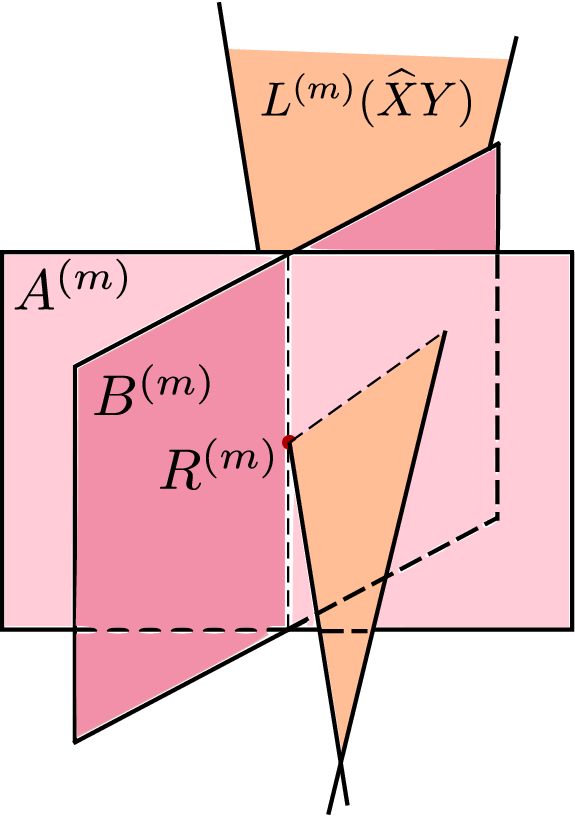}
  }\\
  \includegraphics[width=0.35\columnwidth]{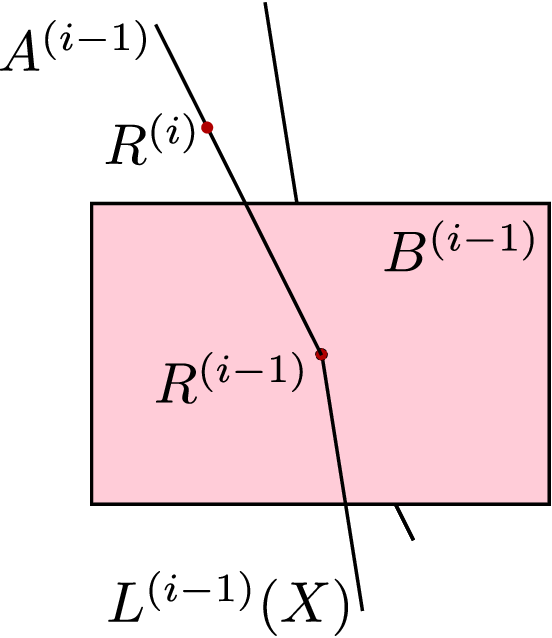}
  \hspace{20pt} \raisebox{30pt}{\Large$\rightarrow$} \hspace{15pt}
  \raisebox{-5pt}{
  \includegraphics[width=0.35\columnwidth]{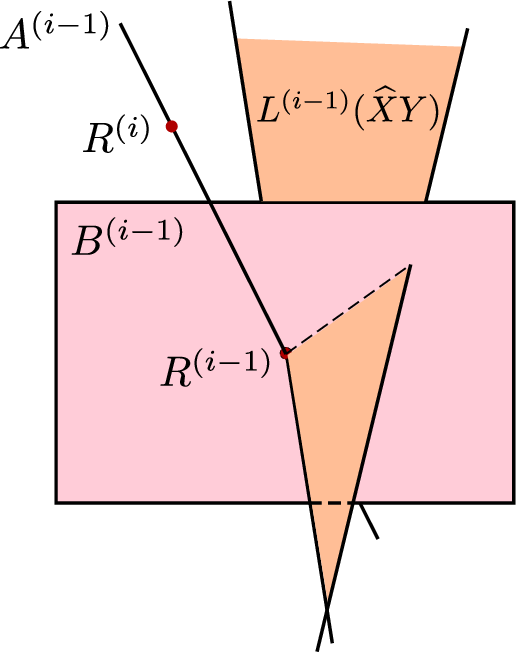}
  }
\caption{The geometric relation indicated by the ground level (the upper figures) and the $i$-th level (the lower figures) of the nested minor. The left figures are the relations before the BCFW shift. The right figures are the relations after the BCFW shift.  }
\label{fig:NestCstrRn}
\end{figure}

To write a  constraint in a compact form and make the   geometric  relations encoded  manifest,
We define a line $L(X)$, for  $i\in [2,m]$
\vspace{-5pt}
$$
L^{(i)}(X)
    \equiv \Big(L^{(i-1)}(X)\cap (B^{(i-1)}_1
    \cdots B^{(i-1)}_{b_{i-1}})A^{(i-1)}_1 \Big).
\vspace{-5pt}
$$
We further define a point $ R^{(i)}(X)\equiv L^{(i)}(X)\cap (B^{(i)}_1\cdots
B^{(i)}_{b_{i}})$, for $i\in [2,m]$.
Given a minor $M(X)$, we could obtain the point $R^{(m)}(X)$ as 
$$
R^{(m)}(X)=L^{(m)}(\widehat{X}Y)\cap(B_1^{(m)}
   \cdots B_{b_m}^{(m)})\cap(A_1^{(m)}\cdots
   \bar A^{(m)}_{\bar a_m}).$$
 All levels of $R^{(i)}(X)$ can be recursively obtained according to 
 $$
R^{(i-1)}= (R^{(i)}A_1^{(i-1)})\cap\left(L^{(i-1)}(\widehat{X}Y)  \cap(B_1^{(i-1)}\cdots B_{b_{i-1}}^{(i-1)})\right).$$
The geometrical relations is shown in Fig.~\ref{fig:NestCstrRn}.
Finally we are able to denote the column $X$ using the columns in the shifted Grassmannian, $X=(R^{(1)}A_1^{(0)})\cap(\widehat{X}Y).$
After removing the constraint $M(X)=0$, the remaining constraints are invariant under the
$\widehat{C}$ representation of $X$, making them independent of the shift $(Y,X)$.

For now we have obtained geometry constraints according to BCFW bridge decomposition chain. We would like to stress that our approach can be applied to seeking all loop leading singularity's geometry constraints. During the process, we introduced a method that the constraints are independently and completely represented. The constraints of the graph constructed by any ``top-form bridge" are immediately obtained using our method. Thus the top-form integrations' contour $\Gamma$ is determined.

\subsection{Rational top-forms  and linear BCFW bridges}
\label{sec:rat}
\vspace{-10pt}

Attaching BCFW bridges  and using the 3- or 4-point amplitude relations reductively,
all non-planar diagrams can be constructed and their $d$log forms be found.
We should stress, however, that not all nonplanar on-shell diagrams have rational top-forms; and it is worth to remark on which kind of non-planar on-shell diagrams  can  have  rational top-forms.
We  address  this question  by building up an equivalent relation between
\textit{rational top-form} and
\textit{linear BCFW bridges}.
If a BCFW bridge results in the shifted  constraint function  to be a linear function of $\alpha$,
 we call this BCFW bridge \textit{a linear BCFW bridge}.

A constraint function $F_i$ is a rational function of the minors of Grassmannian matrix, $C$.
Altogether they span an algebraic ideal $\mathcal{I}[\{F_i\}]$.
Under a BCFW shift $ X \rightarrow \widehat{X} =X + \alpha Y$,  a constraint is eliminated, with
 $C$ being transformed to $\widehat C$.
The transformed $f^{\prime}(\widehat C)$ is also rational iff $\alpha$ is also a rational function of $\widehat C$. 

Next we need to show that rationality of $\alpha$ is guaranteed by the linear BCFW shifts.
To prove their equivalence, assuming
$\alpha={ P(\widehat{C})/Q(\widehat{C}) },$
where $P,\, Q$ are polynomials of minors of
$\widehat C$.
Expanding  $P,\, Q$  as polynomials of $\alpha$ as well as the minors of $C$,
\vspace{-5pt}
$$
\alpha
=\frac{P_0+P_1 \alpha + P_2 \alpha^2 +\cdots +P_N \alpha^N}%
{Q_1  + Q_2 \alpha  + \cdots  +  Q_N \alpha^{N-1} }.
\vspace{-5pt}
$$
The coefficient in each power of $\alpha$, such as $P_0, P_1-Q_1, P_2-Q_2, \cdots, P_N-Q_N$,
is supposed to vanish.
If any of them appears nonzero, it must fall into the ideal $\mathcal{I}[\{F_i\}]$.
This means that  all the coefficients are constraints of $C$.
Under the shift,  $P_i$ and $Q_i$ become
\begin{eqnarray}
\widehat{P}_i &=& P_i + \sum_{j=1}^{N-i} {j\choose i+j} P_{i+j}\alpha^j\nb\\
\widehat{Q}_i &=& Q_i +\sum_{j=1}^{N-1-i} {j\choose i+j} P_{i+j}\alpha^j.
\end{eqnarray}
The constraint $P_{N}-Q_{N}$ remains the same after the  shift
$\widehat{P}_N-\widehat{Q}_N = P_N-Q_N$.
The constraint
$P_{N-1}-Q_{N-1}=0$
appears linearly dependent on  $\alpha$ upon
\vspace{-5pt}
$$
\widehat{P}_{N-1}-\widehat{Q}_{N-1} = [NP_N-(N-1)Q_N]\alpha.
\vspace{-5pt}
$$
If
$[NP_N-(N-1)Q_N]$
 does not vanish then the constraint
$P_{N-1}-Q_{N-1}=0$
 is the removed. Otherwise
$P_N = Q_N = 0$
 and the constraint
$P_{N-2}-Q_{N-2}=0$
 is shifted linearly,
\vspace{-5pt}
$$
\widehat{P}_{N-2}-\widehat{Q}_{N-2}
= [(N-1)P_{N-1}-(N-2)Q_{N-1}]\alpha.
\vspace{-5pt}
$$
Since $P(\widehat{C})$ cannot be totally independent of
$\alpha$, we can trace the constraints from $N$ to $0$ until we find one constraint $(P_i-Q_i)$ which is a linear function of $\alpha$ after a shift.
This constraint is then
the constraint being removed.

The proof of the reverse is also straightforward:
if one constraint becomes  a linear function of
$\alpha$ under a shift, for instance
$$F_i(C) \rightarrow F_i(\widehat{C})
= F_i(C) + F'_i(C)\alpha = F'_i(\widehat{C})\alpha,$$
where $F_i(C)$ vanishes and $F'_i(C)$ is invariant under the shift, we have
 $\alpha = {F_i(\widehat{C})/F'_i(\widehat{C})}.$
Note that the remaining  $s-1$  constraints can be written in the form
$F_{i}(\cdots \widehat{X}-\alpha {Y} \cdots )=0$,
for $i\in[2, s]$,
which are invariant under the shift.

Finally we conclude that upon adding a BCFW bridge
the  on-shell diagram resulted  has a rational top-form if and only if the shift  on the algebra ideal  $\mathcal{I}$ is linear.
For a generic  on-shell diagram, BCFW-bridges can be added in an arbitrary manner and the transformations
on the constraints  are complicated.
Top-forms  can be obtained if and only if  when the BCFW parameters shift the constraints linearly.
This type of bridges is thus called \textit{linear BCFW bridges}.
In the construction of top-forms one should avoid using BCFW bridges that shift the constraints in a nonlinear manner.

\subsection{From BCFW-decompositions to top-forms}
To obtain the top-form of scattering amplitudes, besides the geometric  constraints, we also need to get the integrand, $f(C)$. It must then contain those poles equivalent to the constraints in $\Gamma$ to keep the non-vanishing of the circle-integration in Eq. \ref{eq:topform}.
Each BCFW bridge removes one pole in $f(C)$ by shifting a zero minor to be nonzero: in tangled cases the poles in the integrand must change their forms accordingly.

To see this  we parameterise the constraint  matrix, $C$, using the BCFW parameter, $\alpha$.
In the last BCFW shift
$
X\rightarrow \widehat{X}=X+\alpha Y,
$
several minors in $f(C)$ become functions of $\alpha$.
There exists at least one minor
$
M_0(\widehat{X})=M_0(X)+\alpha R(Y)
$
having  a pole at $\alpha=0$.
After this shift,
$
M_0(X)\rightarrow M_0(\widehat{X}),
$
the constraint $M_0(X)=0$ is removed. And
$
\alpha=M_0(\widehat{X})/R(Y)
$
is then a rational function of $\widehat{C}$
and can be subtracted from  other  shifted minors to obtain the shift-invariant  minors of $\widehat{C}$,
$
M_i(X)=M_i(\widehat{X}-\alpha Y).
$
This is demonstrated in Sec.~\ref{sec:rat}

We can further attach a  BCFW bridge to the integrand, %
\begin{equation}
\label{Topfc}
  f(\widehat{C})
= M_0(\widehat{X}) \prod_i M_i(\widehat{X}-\alpha Y)
\times
\left(
   \begin{array}{c}
   \text{minors}\\ \text{without}~\alpha
   \end{array}
\right).
\end{equation}
In this way top-forms of leading singularities of scattering amplitudes can thus
be obtained--be it planar or nonplanar--and from tree level to all loops.
We  illustrate our method below with several examples:  searching for  the constraints and calculating the top-form integrand.

\subsection{Several Examples}
\begin{figure}[htbp]
  \centering
  \includegraphics[width=0.5\textwidth]{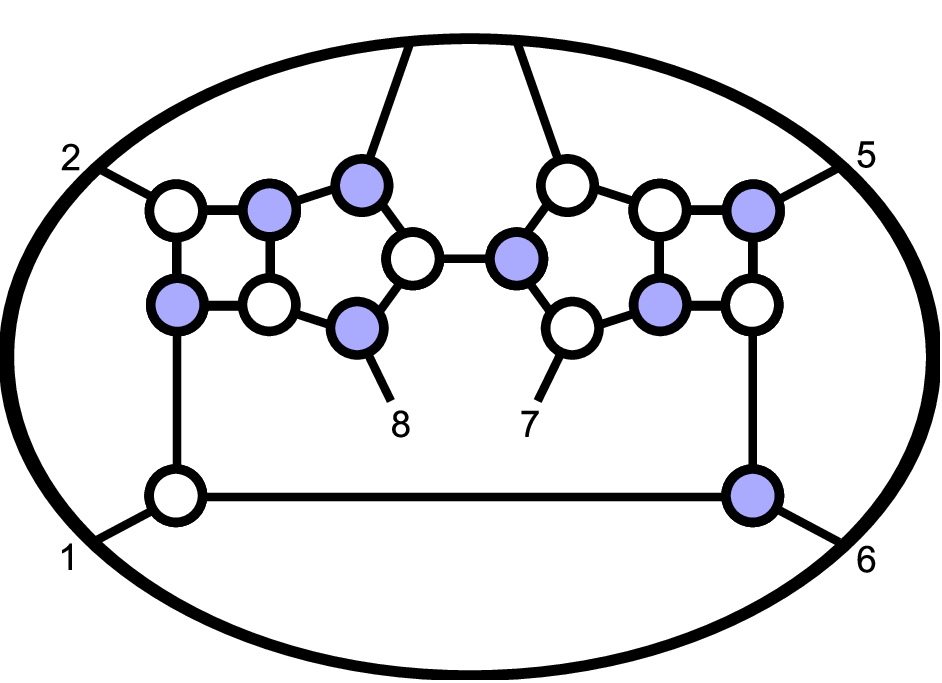}
  \caption{A nonplanar on-shell  diagram of $\mathcal{A}_{8}^{4}$}\label{fig:A84topform}
\end{figure}

\paragraph{A one loop example of attaching a nonadjacent bridge}~
As an application, we take the nonplanar diagram in Fig.~\ref{fig:A84topform} as an example.
According to the permutation of the planar diagrams before attaching the bridge $(1,6)$, linear relations of the diagram are $(4,5,6,7)^{2}$ and $(8,1,2,3)^{3}$. And the top-form is 
\begin{equation}
\mathcal{T}_{8}^{4}=\oint\limits_{\Gamma}\frac{d^{16}C}{(1234)(2345)(3456)(4567)(5678)(6781)(7812)(8123)},
\end{equation}
where we have omitted the delta functions for clarity. Attaching bridge $(1,6)$, the coplanar constraint  $(8,1,2,3)^{3}$ is unaffected  while $(4,5,6,7)^{2}\rightarrow (4,5,7)^2, ((4,5,7)^21\hat 6)^3$, where $\hat 6=6+\alpha 1$. We choose the shifted pole as  $(3456)$. The shift parameter $\alpha$ can be written as  $\alpha=\frac{(345\hat 6)}{(3451)}$. According to Eq. \ref{Topfc}, we get
\begin{equation}
\mathcal{T}_{8}^{4}=\oint\limits_{\Gamma}\frac{d^{16}C}{(1234)(2345)(3456)(45(\hat 6 -\alpha 1)7)(5(\hat 6-\alpha 1)78)(6781)(7812)(8123)},
\end{equation}
where $$(45(\hat 6 -\alpha 1)7)=(45\hat 6 7)-\frac{(345\hat 6)}{(3451)}(4517)=-{(3457)(1456)\over (3451)}$$ and $$(5(\hat 6 -\alpha 1)78)=(5\hat 6 78)-\frac{(345\hat 6)}{(3451)}(5178)=-{(3458)(1567)\over (3451)}.$$
Finally we obtain top-form of $\mathcal{A}_{8}^{4}$:
$$\widehat{\mathcal{T}_{8}^{4}}=\oint\limits_{\widehat{\Gamma}}\frac{d^{4\times8}\widehat{C}}{vol(GL(4))}\frac{(1345)^2}{(1234)(2345)(3456)(3457)(1456)(3458)(1567)(6781)(7812)(8123)}.$$

Therefore, we can always construct the top-from of nonplanar diagrams by attaching adjacent and nonadjacent bridges on identity diagram step by step.

\paragraph{A tangled  two-loop example}
At  multi-loop level the geometric constraints for a nonplanar leading singularity can be highly tangled, as the diagrams cannot, in general,  be reduced to the planar ones by KK-relation~\cite{KK}.

The diagram is a planar one before attaching the bridge (3,5) in the seventh step and the top-form is
\begin{equation}
\mathcal{T}_{6}^{3}=\oint\limits_{\Gamma}\frac{d^{9}C}{(123)(234)(345)(456)(561)(612)},\nb
\end{equation}
where we have omitted the delta functions for clarity. After attaching BCFW bridge $(3,5)$, the on-shell diagram become planar. The top-form can be obtained directly as in last paragraph
\begin{equation}
\mathcal{T}_{6}^{3}=\oint\limits_{\Gamma}\frac{d^{9}C (361)}{(123)(234)(345)(356)(146)(561)(612)},
\end{equation}
here the constraints $\Gamma$ is determined by the linear relation $(2,3,4)^2$ and  $(3,5,6)^{2}$ as shown in Table \ref{TwoLoopGeo}. We do not distinguish the labels $\Gamma$ for each step in adding the BCFW bridges.
According the discussion above, upon attaching the last BCFW bridge $(1,3)$. Its exact expression can be obtained by transforming the constraints in the last step linearly as
$ (234)=(356)=0   \rightarrow (234)=0,
\frac{(234)}{(214)}-\frac{(356)}{(156)}=0.
 $
Adding the bridge $(1,3)$ and the elimination of $(234)=0$ leaves
$\frac{(234)}{(214)}-\frac{(356)}{(156)}=0$
invariant. Applying the same linear transformation to the denominator:
$$
 \frac{1}{ (234) (356)}
\rightarrow
\frac{-1}{(2\widehat 34)\left(\frac{-(156)(2\widehat 34)}{(214)}+(\widehat 356)\right)}~,
$$
we extract  the top-form
\begin{eqnarray}  \label{eq:warmup}
\widehat{\mathcal{T}_{6}^{3}}
=\oint\limits_{\Gamma}
\frac{d^{9}\widehat{C}(214)^{2}(\widehat 361)}{(12\widehat 3)(2\widehat 34)\left((1\widehat 3)\cap(24)56\right)\left((1\widehat 3)\cap(24)45\right)(146)(561)(612)}.\nonumber
\end{eqnarray}

\paragraph{MHV top-form and its simplification}The MHV top-form can be further simplified. We can always transform any nonplanar MHV top-form into a summation of several top-forms. These generated top-forms share features that their numerators of the integrands equal 1 and the minors in $f(C)$ are of cyclic orders, which are exactly those of planar MHV top-forms. This yields a strong proof that any nonplanar MHV amplitude is a summation of several planar amplitudes.

To see this, let us consider the top-form of $\mathcal{A}_{n}^{2}$. Attaching a nonadjacent bridge $(a,b)$ to a planar diagram yields
$$\widehat{\mathcal{T}_{n}^{2}}=\int\frac{d\alpha}{\alpha}\oint\limits_{\Gamma}\frac{d^{2\times n}C}{vol(GL(2))}\frac{\delta(C)}{(12)\cdots(b-1,b)(b,b+1)\cdots(n1)}$$
Without losing generality, we assume the pole at $(b-1,b)=0$. Following the same procedure illustrated in Eq. \ref{eq:topform}, we obtain
$$\frac{1}{f(\widehat{C})}=\frac{(a,b-1)}{(12)\cdots(b-2,b-1)(b-1,\widehat{b})(a,\widehat{b})(b+1,b-1)(b+1,b+2)\cdots(n1)}$$

Since $a<b-1$, we define $a+m=b-1$ $(m\in\mathbb{Z}^{+})$.
\begin{itemize}
\item If $m=1$, the numerator is then $(b-2,b-1)$ and the integrand can be simplified to a term with its numerator equaling one and $f(C)$ of cyclic orders, i.e. a planar one.
\item If $m>1$ , we can multiply the numerator and denominator by $(a+1,\widehat{b})$:
$$\begin{array}{rcl}
    \frac{1}{f(C)} & = & \frac{(a,b-1)(a+1,\widehat{b})}{(12)\cdots(b-2,b-1)(b-1,\widehat{b})(a,\widehat{b})(b+1,b-1)(b+1,b+2)\cdots(n1)(a+1,\widehat{b})} \\
     & = & \frac{(a,a+1)(\widehat{b},b-1)+(a,\widehat{b})(a+1,b-1)}{(12)\cdots(b-2,b-1)(b-1,\widehat{b})(a,\widehat{b})(b+1,b-1)(b+1,b+2)\cdots(n1)(a+1,\widehat{b})} \\
     & = & \frac{1}{(12)\cdots(a-1,a)(a+1,a+2)\cdots(b-2,b-1)(a,\widehat{b})(b-1,b+1)(b+1)(b+2)\cdots(n1)(a+1,\widehat{b})}+\\
     & &\frac{(a+1,b-1)}{(12)\cdots(b-2,b-1)(b-1,\widehat{b})(a+1,\widehat{b})(b+1,b-1)(b+1,b+2)\cdots(n1)}.
  \end{array}
$$
The first term is already planar,  while the second is not obvious.
\begin{itemize}
\item If $m=2$, the second term is planar.
\item If $m>2$, we multiply the integrand by $(a+2,b),(a+3,b),\cdots,(a+m-1,b)$ one by one. For each step of multiplication, we utilize the Pluck relation to transform the nonplanar term into a summation of planar terms and a remaining term. The final term left after series of multiplication is
$$\frac{(a+m-1,b-1)}{(12)\cdots(b-2,b-1)(b-1,\widehat{b})(a+m-1,\widehat{b})(b+1,b-1)(b+1,b+2)\cdots(n1)}.$$
Since $a+m=b-1$, this term is also planar.
\end{itemize}
\end{itemize}
Following these steps, we can finally simplify the top-forms of all nonplanar MHV amplitudes into the sum of planar ones. One can easily verify that the simplification process from nonplanar one to planar term's summation is equivalent to applying KK relation to MHV amplitudes.

In this section, we construct  the top-forms of the nonplanar on-shell graphs.
The key step  is attaching a nonadjacent BCFW bridge to a planar diagram.
The cyclic order of $f(C)$ is then  broken  and we obtain a different integrand from the planar ones.
Keep attaching bridges on the  identity and we can  arrive at the top-form of our target--the nonplanar leading singularity.
We  then break down  the top-forms of the nonplanar MHV amplitudes into a  summation of the planar top-forms.
For the leading singularities of the  one-loop amplitudes, this  simplification is similar to the  KK relation.
For leading singularities of the  general amplitudes, the relation between the  top-form's simplification
and the  KK relation will be discussed in our future   work.



\section{conclusion}
We  have classified  nonplanar on-shell diagrams 
according to whether they posses rational
top-forms, and proved its equivalence to linear BCFW bridges. 
We conclude that when attaching linear bridges, geometric constraints of the nonplanar diagrams--tangled or untangled--%
can all be constructed systematically.
With this  chain of BCFW bridges rational top-forms of the nonplanar on-shell diagrams can then be derived  in a straightforward way.  
This method applies to  leading singularities of 
nonplanar multi-loop amplitudes beyond MHV.


\begin{acknowledgements}
GC thanks Nima Arkani-Hamed for helpful discussion and useful comments. We thank Peizhi Du, Shuyi Li and Hanqing Liu for constructive discussion. Yuan Xin thanks Bo Feng for introducing the background on the recent developments of scattering amplitude.  GC, RX and HZ have been supported by the Fundamental Research Funds for the Central Universities under contract 020414340080, NSF of China Grant under contract 11405084, the Open Project Program of State Key Laboratory of Theoretical Physics, Institute of Theoretical Physics, Chinese Academy of Sciences, China (No.Y5KF171CJ1). We also thank Y. Gao, T. Han for hospitality and Key Laboratory of Theoretical Physics for hosting. 
\end{acknowledgements}

\section{Appendix}
\label{sec:appendix}
All rational top-forms can be constructed by our method described above.
In our discussion we have assumed that the BCFW bridges are the
\textit{linear BCFW bridges}--each successive
$\alpha$-shift transforms the constraints linearly
and can thus be represented by a rational function of minors of the underlying constraint C-matrix.
However not all on-shell diagrams are  made up
completely of such bridges:
the constraints can be nonlinear in $alpha$ and
cannot be written as rational functions under some shift.
 Such an on-shell diagram will not have a rational top-form.

We present a counter example, $\mathcal{A}_{10}^{4}$.
Upon attaching the bridge $(3,4)$,
two constraints $(2,3,4,8)^{3}$ and $(1,3,4,6)^{3}$
emerge (the superscripts denote the number of independent column vectors), with two columns, $3$ and $4$, being the same.
Attaching  the bridge $(5,3)$, one of the constraints is removed and the other one becomes
$((5,3)\cap(2,4,8),1,4,6)^{3}$.
If we go on attaching  the bridge $(7,4)$, the tangled constraint is removed.
However, due to the column $4$ appearing twice in that constraint, such a bridge results in a nonlinear shift of the algebraic ideal.
Therefore such a $\alpha$-shift cannot be represented linearly by some minor being zero,  violating our linearity requirements in the construction of rational top-forms.
\begin{figure}[htbp]
  \centering
  \includegraphics[width=0.45\textwidth]{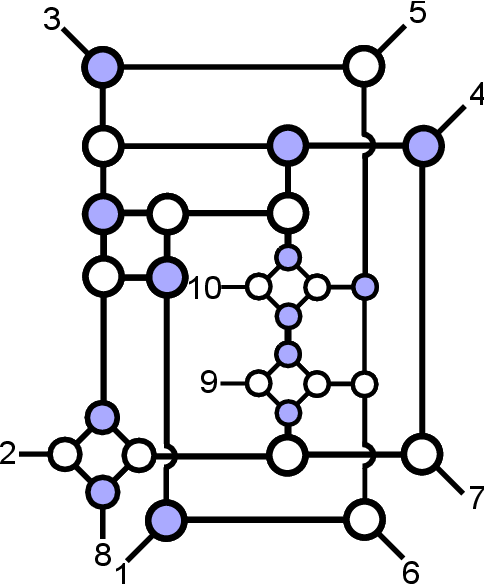}
  \caption{A nonplanar diagram of $\mathcal{A}_{10}^{4}$}\label{fig:A104}
  \label{ConterEx}
\end{figure}
\begin{table}[htb]
\caption{The evolution of the geometric constraints with adding BCFW bridges for the diagram in Fig.~\ref{ConterEx}.}
\begin{center}
\begin{tabular}{|c|cccccc|}\hline
&$(3)^{0}$&$(4)^{0}$&$(7)^{0}$&$(8)^{0}$&$(9)^{0}$&$(10)^{0}$\\ \hline 
$(6,10)$&$(3)^{0}$&$(4)^{0}$&$(7)^{0}$&$(8)^{0}$&$(9)^{0}$&$(6,10)^{1}$\\ \hline 
$(2,6)$&$(3)^{0}$&$(4)^{0}$&$(7)^{0}$&$(8)^{0}$&$(9)^{0}$&$(2,6,10)^{2}$\\  \hline 
$(6,9)$&$(3)^{0}$&$(4)^{0}$&$(7)^{0}$&$(8)^{0}$&$(6,9)^{1}$&$(2,6,10)^{2}$\\ \hline 
$(5,6)$&$(3)^{0}$&$(4)^{0}$&$(7)^{0}$&$(8)^{0}$&$(5,6,9)^{2}$&$(2,6,10)^{2}$\\ \hline 
$(6,9)$&$(3)^{0}$&$(4)^{0}$&$(7)^{0}$&$(8)^{0}$&$(5,6,9)^{2}$&$(2,6,9,10)^{3}$\\ \hline 
$(6,8)$&$(3)^{0}$&$(4)^{0}$&$(7)^{0}$&$(6,8)^{1}$&$(5,6,9)^{2}$&$(2,6,9,10)^{3}$\\ \hline 
$(6,7)$&$(3)^{0}$&$(4)^{0}$&$(6,7)^{1}$&$(6,8)^{1}$&$(5,6,9)^{2}$&$(2,6,9,10)^{3}$\\ \hline 
$(5,6)$&$(3)^{0}$&$(4)^{0}$&$(7,8)^{1}$&$(5,6,8)^{2}$&$(5,8,9)^{2}$&$(2,8,9,10)^{3}$\\ \hline 
$(2,5)$&$(3)^{0}$&$(4)^{0}$&$(7,8)^{1}$&$(6,8,9)^{2}$&$(2,5,8,9)^{3}$&$(2,8,9,10)^{3}$\\ \hline 
$(2,4)$&$(3)^{0}$&$(2,4)^{1}$&$(7,8)^{1}$&$(6,8,9)^{2}$&$(2,5,8,9)^{3}$&$(2,8,9,10)^{3}$\\ \hline 
$(1,2)$&$(3)^{0}$&$(1,2,4)^{2}$&$(7,8)^{1}$&$(6,8,9)^{2}$&$(4,5,8,9)^{3}$&$(4,8,9,10)^{3}$\\ \hline 
$(2,3)$&$(2,3)^{1}$&$(1,2,4)^{2}$&$(7,8)^{1}$&$(6,8,9)^{2}$&$(4,5,8,9)^{3}$&$(4,8,9,10)^{3}$\\ \hline 
$(1,2)$&$(1,2,3)^{2}$&$(1,3,4)^{2}$&$(7,8)^{1}$&$(6,8,9)^{2}$&$(4,5,8,9)^{3}$&$(4,8,9,10)^{3}$\\ \hline 
$(8,2)$&$(1,2,3,8)^{3}$&$(1,3,4)^{2}$&$(7,8)^{1}$&$(6,8,9)^{2}$&$(4,5,8,9)^{3}$&$(4,8,9,10)^{3}$\\ \hline 
$(2,8)$&$(1,2,3,8)^{3}$&$(1,3,4)^{2}$&$(2,7,8)^{2}$&$(6,7,9)^{2}$&$(4,5,7,9)^{3}$&$(4,7,9,10)^{3}$\\ \hline 
$(6,1)$&$(2,3,4,8)^{3}$&$(1,3,4,6)^{3}$&$(2,7,8)^{2}$&$(6,7,9)^{2}$&$(4,5,7,9)^{3}$&$(4,7,9,10)^{3}$\\ \hline 
$(3,4)$&$(2,3,4,8)^{3}$&$(1,3,4,6)^{3}$&$(2,7,8)^{2}$&$(6,7,9)^{2}$&$(5,7,9,10)^{3}$&\\ \hline 
$(5,3)$&$ $&$((5,3)\cap(2,4,8),\atop 1,4,6)^{3}$&$(2,7,8)^{2}$&$(6,7,9)^{2}$&$(5,7,9,10)^{3}$&\\ \hline 
$(7,4)$&$ $&$ $&$(2,7,8)^{2}$&$(6,7,9)^{2}$&$(5,7,9,10)^{3}$&\\ \hline 
\end{tabular}
\end{center}
\label{default}
\end{table}

\end{document}